\documentclass[pdflatex,sn-mathphys-num]{sn-jnl}

\usepackage{graphicx}%
\usepackage{multirow}%
\usepackage{amsmath,amssymb,amsfonts}%
\usepackage{amsthm}%
\usepackage{mathrsfs}%
\usepackage[title]{appendix}%
\usepackage{xcolor}%
\definecolor{medicinecluster}{RGB}{128, 203, 179}
\definecolor{cscluster}{RGB}{247, 157, 122}
\usepackage{textcomp}%
\usepackage{manyfoot}%
\usepackage{booktabs}%
\usepackage{algorithm}%
\usepackage{algorithmicx}%
\usepackage{algpseudocode}%
\usepackage{listings}%
\usepackage{subcaption}%
\usepackage{tcolorbox}%
\usepackage{float}
\usepackage{lineno}

\usepackage{geometry}
\theoremstyle{thmstyleone}%
\theoremstyle{thmstyletwo}%

\theoremstyle{thmstylethree}%

\begin{document}

\title[Article Title]{Academic collaboration on large language model studies increases overall but varies across disciplines}


\author*[1]{\fnm{Lingyao} \sur{Li}}\email{lingyaol@umich.edu}
\equalcont{These two authors contributed equally to this work.}
\author*[2]{\fnm{Ly} \sur{Dinh}}\email{lydinh@usf.edu}
\equalcont{These two authors contributed equally to this work.}
\author[3]{\fnm{Songhua} \sur{Hu}}\email{hsonghua@mit.edu}

\author[1]{\fnm{Libby} \sur{Hemphill}}\email{libbyh@umich.edu}

\affil*[1]{\orgdiv{School of Information}, \orgname{University of Michigan}, \orgaddress{\city{Ann Arbor}, \state{MI}, \country{United States}}}

\affil[2]{\orgdiv{School of Information}, \orgname{University of South Florida}, \orgaddress{\city{Tampa}, \state{FL}, \country{United States}}}

\affil[3]{\orgdiv{Senseable City Laboratory}, \orgname{Massachusetts Institute of Technology}, \orgaddress{\city{Cambridge}, \state{MA}, \country{United States}}}


\abstract{Interdisciplinary collaboration is crucial for addressing complex scientific challenges. Recent advancements in large language models (LLMs) have shown significant potential in benefiting researchers across various fields. To explore the application of LLMs in scientific disciplines and their potential for interdisciplinary collaboration, we collect and analyze data from OpenAlex, an open-source academic database. Our dataset comprises 59,293 LLM-related papers, along with two control groups---a random sample of 70,945 machine learning (ML) papers and 73,110 papers from non-LLM/ML fields. We first employ Shannon Entropy to assess the diversity of collaboration in terms of authors' institutions and departments. Our results reveal that many fields have exhibited a more significant increasing trend in entropy following the release of ChatGPT as compared to the control paper groups. In particular, \textit{Computer Science} and \textit{Social Science} display a consistent increase in both institution and department entropy. Other fields such as \textit{Decision Science}, \textit{Psychology}, and \textit{Health Professions} have shown minor to significant increases in 2024 compared to 2023. Our difference-in-difference analysis also indicates that the release of ChatGPT leads to a statistically significant increase in collaboration in several fields, such as \textit{Computer Science} and \textit{Social Science}. In addition, we analyze the author networks and find that \textit{Computer Science}, \textit{Medicine}, and other \textit{Computer Science}-related departments are the most prominent in LLM research. Regarding authors' institutions, our analysis reveals that entities such as \textit{Stanford University}, \textit{Harvard University}, and \textit{University College London} are key players, either dominating centrality measures or playing crucial roles in connecting research networks. Overall, this study provides valuable information on the current landscape and evolving dynamics of collaboration networks in LLM research. It also suggests potential areas for fostering more diverse collaborations and highlights the need for continued research on the impact of LLMs on scientific practices and outcomes.}


\maketitle


\section{Introduction}
\label{sec1}

Interdisciplinary collaboration is crucial for tackling complex scientific challenges, involving investigators from various fields of expertise \cite{ShiEvans2023SurprisingCombos,dinh2024hyperauthored,venturini_collaboration_2024}. Recent advances in generative AI models and applications, particularly the development of large language models (LLM) such as OpenAI's ChatGPT, Google's PaLM, and Meta's Llama, impact research activities across disciplines. These models find applications in research tasks such as literature search \cite{khraisha2024can,le2023chatgpt}, content analysis \cite{pilny2024manual}, and findings generation from provided data \cite{byun2023dispensing}. The integration of LLMs into research workflows has enabled researchers to process and analyze vast amounts of scientific data more efficiently, identify patterns and connections across disciplines, and generate preliminary insights that can guide further investigation.

While researchers initially developed and applied LLMs within computer science, their use has expanded into other disciplines, leading to an increasing number of studies in areas such as medicine, social sciences, and business. In health-related fields, LLMs are being employed to interpret protein structures \cite{ferruz2022protgpt2}, process electronic health records \cite{yang2022large}, and even aid in drug discovery \cite{savage2023drug}. Engineering benefits from these models through advancements in autonomous driving \cite{ma2024lampilot} and remote sensing \cite{kuckreja2024geochat} technologies. Social scientists leverage LLMs for large-scale text analysis, including attitude simulation \cite{argyle2023out} and content moderation on social media platforms \cite{li2024hot}. Their impact also extends to finance, where professionals use LLMs to streamline document review and perform financial analysis \cite{wu2023bloomberggpt}. The wide-ranging applicability of LLMs across disciplines suggests that they may influence many aspects of the research process, including how researchers collaborate. 

Traditionally, interdisciplinary collaboration has played a crucial role in combining expertise from various fields to discover innovative solutions to scientific challenges. Researchers increasingly use tools like ChatGPT to gain cross-disciplinary insights and interpret their work through different domains and perspectives \cite{susarla2023janus}. Researchers also use LLMs to automate certain aspects of the scholarly writing process. They synthesize existing literature from multiple disciplines on a specific topic \cite{dwivedi2023opinion} and reframe the main findings into domain-specific language \cite{einarsson2024application}. 

By automating routine research tasks \cite{le2023chatgpt,owens2023nature} associated with manuscript formatting or data entry, LLMs can help free up time for researchers to build their collaborative networks and deepen their engagement in cross-disciplinary literature \cite{agathokleous2023use,dwivedi2023opinion}. Well-known LLMs such as ChatGPT have also illustrated capabilities to bridge language gaps between disciplines such as environmental science and ecology \cite{dwivedi2023opinion}, statistics and biology \cite{einarsson2024application}. Prior analysis of ChatGPT's use in environmental sciences shows that LLMs may improve shared language between researchers \cite{agathokleous2023use}, which in turn facilitates shared understanding across different domains. However, it's important to recognize that LLMs' effectiveness is primarily illustrated in language-related tasks. LLMs are not substitutes for the critical examination of findings or the nuanced and domain-specific understanding essential for truly bridging disciplinary language \cite{meyer2023chatgpt,barleydinh2022interdisnets}.


Scientific collaboration has long been a compelling research topic within the academic community \cite{fortunato2018science}. Previous studies have primarily used network analysis to characterize these collaborations \cite{coccia2016evolution, alshebli2018preeminence} or explored their effects on the quality and novelty of scientific outputs \cite{shin2022scientific, lariviere2015team}. Based on the evidence that LLMs are beginning to impact the way scholars work and collaborate, the questions then arise: Whether and how have LLMs transformed interdisciplinary collaboration? In what ways have they changed how different disciplines interact and collaborate? While a few recent studies have leveraged bibliometric analysis to examine scientific collaboration in the context of LLMs \cite{fan2024bibliometric} or their applications in specific fields such as biomedical and health sciences \cite{li2024scoping}, they do not address how LLMs have transformed interdisciplinary collaboration. In particular, given the general increase in team sizes and the rise of interdisciplinary research, as highlighted in prior studies \cite{khabsa2014number, bornmann2015growth}, demonstrating the unique impact of LLMs on scientific collaboration presents an additional challenge.

To address the research gaps, our study explores the application of LLMs in scientific disciplines and their implications for interdisciplinary collaboration by addressing two research questions, as listed below. The first question examines the diversity of co-authors in terms of their institutional and departmental affiliations, using the Shannon Entropy measure. The second question aims to understand the structural patterns of co-authorship collaborations, using network analysis to identify key institutions and departments that are active producers of LLMs research, as well as those facilitating collaborations across disciplinary boundaries. To isolate the impact of LLMs from broader trends in scientific research, such as the general increase in team sizes or variations in productivity across different disciplines \cite{bornmann2015growth,khabsa2014number,tabah1999literature,wuchty2007increasing}, we compare the impact of LLMs against two control groups: papers from machine learning (ML papers) and papers that do not involve either LLM or ML (non-LLM/ML papers). This comparative approach allows us to assess whether the observed changes in collaboration patterns are attributable to the use of LLMs or if they are part of broader shifts occurring across all scientific disciplines. 

\begin{itemize}
    \item \textbf{RQ1}. How diverse are the coauthors of papers utilizing LLMs in terms of their research institutions and departments across different fields, in comparison to the control groups? 
    \item \textbf{RQ2}. What are the structural patterns of co-authorship networks in research utilizing LLMs, and what roles do key entities (leading institutions and departments) play in facilitating and enhancing collaboration, in comparison to the control groups?
\end{itemize}

To answer these questions, we collect 59,293 papers from OpenAlex related to LLMs from 27 scientific fields published between the release of the BERT model in October 2018 and September 2024. We use Shannon Entropy to measure the authorship diversity and network techniques to reveal the collaboration structures. Our findings indicate that since the advent of LLMs, particularly after the release of ChatGPT in November 2022, collaboration diversity has increased across many fields, notably \textit{Computer Science} and \textit{Social Science}. Overall, our findings demonstrate that LLM research has grown exponentially since 2022 and holds the potential to enhance interdisciplinary collaboration. It also encourages the involvement of prominent academic institutions in leading LLM-based research and applications in different domains.



\section{Data and Methods}
\label{sec2}

Aligned with the two research questions, our data collection and methods involve two units of analysis: papers and authors. The first research question examines papers to analyze the diversity of coauthors across research fields and institutions. The second research question centers on authors to explore the structural patterns of co-authorship networks and identify key researchers and their roles in facilitating collaboration. Combining these two units of analysis allows us to capture (1) which disciplines and institutions are most impacted by the advent of LLMs in terms of collaboration diversity, as well as (2) key researchers in LLM research and their respective departmental and institutional affiliations. These analyses help us understand how LLMs are impacting scientific collaboration by revealing patterns of interdisciplinary and cross-institutional partnerships, as well as identifying the influential actors driving LLM research across disciplinary boundaries. For each of the two sets of measures, we compare the results against two control groups (ML and non-LLM papers) to ensure the observed trends are specific to LLM researchand not reflective of broader patterns in other AI-related or unrelated fields.

\subsection{Data preparation}
We select OpenAlex \cite{priem2022openalex} for collecting related studies for two main reasons. First, OpenAlex is an open-source repository of scholarly metadata, allowing us to gather both recently archived preprints and published articles from journals and conferences. This is particularly useful for our analysis, given the prevalence of preprinted studies in the field of LLMs. Second, OpenAlex provides its data freely and openly, ensuring that our analysis can be easily replicated by the community without any licensing restrictions.

The workflow for data cleaning is presented in \autoref{workflow}. Within OpenAlex, we collect relevant papers on the most popular LLMs and their respective models, as detailed in \autoref{tab:llm_search_terms} in \autoref{search}. We use two general terms, ``large language model" and ``LLM," along with popular open-source models (e.g., BERT, Flan-T5, LLaMA) and closed-source models (e.g., ChatGPT, Claude) based on the MMLU benchmark \cite{hendrycks2020measuring}. We observe that some models, like Yi or Phi, do not yield relevant papers, potentially introducing significant noise during paper screening. Additionally, models like grok-1 or Galactic did not return any search results. Our data collection results in a total of 177,462 papers.

To ensure the collected papers are relevant to the topic of LLMs, we restrict our search to titles and abstracts. However, some papers containing these keywords might still be irrelevant. Therefore, we implement several steps to filter out irrelevant papers. First, we consider only articles and preprints, excluding types such as editorials and opinions. Second, we remove potential duplicates, including those with duplicated titles and papers initially published as preprints and later as journal articles with similar titles. We use Jaccard similarity to identify and remove these duplicates (see \autoref{jaccard}). Third, we employ GPT-4o mini models to evaluate the relevance of a paper to LLM topics based on its title, abstract, and keywords, and filter out those papers that are not relevant to LLM topics (see \autoref{relevance}). Applying these filtering criteria reduces the original collection to 59,293 papers for the subsequent analysis.

As mentioned earlier, an important question is whether the observed trend could be unique to LLM research or simply reflects a natural progression in academic collaboration. To investigate this, we establish two control groups. The first control group focuses on machine learning (ML) papers. We select ML as a control because it is a well-established field from which LLM emerged as a subfield. To construct this group, we collect a random sampling of 70,945 papers containing the phrase ``machine learning'' in either their title or abstract. Rather than searching for specific ML models like random forest or logistic regression, we use this broader criterion to gather a comprehensive sample while maintaining the simplicity and clarity of resulting dataset. To provide an even broader perspective beyond AI-related fields, we create a second control group consisting of a random sample of 73,110 papers from all other research categories---specifically, papers that belong neither to the ML nor LLM categories. This selection allows us to analyze collaboration patterns in the broader scientific community. The development of control groups is specifically documented in \autoref{workflow}.

\subsection{Measure of collaboration diversity}

Our first research question seeks to assess the diversity of collaboration. OpenAlex offers a variety of information about author affiliations, including details about departments, institutions, and countries. This allows us to represent the authors' affiliation information for a paper using a set as follows,

\begin{equation}
A(x_i) = \{ D(x_i), I(x_i), C(x_i) \} \label{eq1}
\end{equation}

where \(x_i\) denotes the \(i\)$^{th}$ author of a paper, \(A(x_i)\) denotes the set of authors' affiliation information given a paper, \(D(x_i)\) represents their department information, \(I(x_i)\) represents their institution information, and \(C(x_i)\) represents their country information. It is important to note that these three types of information can vary significantly within a single paper. For instance, all collaborating authors could belong to the same institution and country but be affiliated with different departments. Our subsequent analysis particularly focuses on the collaboration between institutions and departments; therefore, we consider the first two sets of variables in \autoref{eq1}.

Next, we use Shannon Entropy to measure the collaboration diversity given authors' affiliation information in a paper. Shannon Entropy quantifies the uncertainty or randomness in a set of possible outcomes. In the context of information theory, it represents the average amount of information produced by a stochastic set of data sources. Mathematically, for a discrete random variable \(Y\) with possible values \(y_1, y_2, \ldots, y_n\) and probability distribution \(P(Y) = \{p(y_1), p(y_2), \ldots, p(y_n)\}\), the Shannon Entropy \(H(Y)\) is calculated as:

\begin{equation}
H(Y) = -\sum_{i=1}^{n} p(y_i) \log_2 p(y_i) \label{eq2}
\end{equation}

where \(Y\) belongs to one of the aspects (e.g., \(D\), \(I\)) in the set of authors' affiliation, \(p(y_i)\) is the probability of the outcome \(y_i\). We use this metric to measure the diversity given authors' affiliation information, such as their affiliated departments. For example, if a paper has five authors with affiliated departments \(D(x_i)\) represented as \(\{d_1(x_1), d_1(x_2), d_2(x_3), d_2(x_4), d_4(x_5)\}\), then the probabilities of \(d_1\), \(d_2\), and \(d_3\) are calculated as 0.4, 0.4, and 0.2, respectively. Using \autoref{eq2}, the Shannon Entropy is: \(H(D) = - \left( 0.4 \log_2 0.4 + 0.4 \log_2 0.4 + 0.2 \log_2 0.2 \right) = 1.5219\). In general, higher entropy indicates greater diversity in collaboration based on authors' affiliation information, while lower entropy suggests that authors' affiliations are more uniform. An entropy of \(0\) implies that all authors of a paper are from the same institution.

While the entropy meassure provides valuable insights into collaboration diversity patterns over time, we are also interested in understanding how the launch of ChatGPT might have affected the scientific collaboration diversity. However, the entropy changes after the launch of ChatGPT cannot be simply attributed to causal effects due to the inherent trends in entropy over time. To further investigate the impact of ChatGPT on the collaboration diversity, a Difference-in-Difference (DiD) model is fitted for each field \cite{brodersen2015inferring, hu2021left} using the following equation:

\begin{equation}
    H(Y) = \beta_0 + \beta_1 \cdot \text{Treatment} + \beta_2 \cdot \text{Post} + \beta_3 \cdot (\text{Treatment} \times \text{Post}) + e
\end{equation}

where $H(Y)$ is the entropy given the time; Treatment is a dummy variable indicating the treatment group (LLM); Post is a dummy variable indicating pre (=0) and post (=1) the launch of ChatGPT; Treatment $\times$ Post is a dummy variable indicating whether the outcome is observed in the treatment group and after the intervention. In addition, $\beta_3$ represents the change in the average entropy of the treatment group after the launch of ChatGPT, compared to what would have been expected in the absence of ChatGPT.  

\subsection{Measures of network structure and network comparison}
We construct co-authorship networks for the LLM papers group, ML group, and non-LLM/ML group, based on the bipartite network projection, which involves converting a paper-author network to a co-authorship network whereby two authors are connected if they have co-authored at least one paper together. Each connection is weighted based on the total number of papers that each pair of researchers co-authored together. This method, as described by \cite{breiger1974duality,dinh2024hyperauthored}, allows us to identify key researchers that have notable collaborative influence in the field, as well as any differences in collaboration patterns depending on the authors' disciplines, institutions, and countries. The formula for our weighted projection approach is shown below:

\begin{equation}
\label{eqn:full_counting}
w_{ij} = \sum a_{i}^{p}a_{j}^{p}
\end{equation}
where $a_{i}^{p}$ denotes whether author $i$ contributes to paper \textit{p} (with 1 indicating authorship and 0 indicating no authorship), and $ a_{j}^{p}$ similarly indicates whether author $j$ contributes to paper \textit{p}. $w_{ij}$ is 1 if $i$ and $j$ are authors of paper \textit{p}. This method assigns a full weight of 1 to each co-authorship instance and sums these weights across all papers where $i$ and $j$ are co-authors.

With the resulting co-authorship networks, we analyze the structural properties in terms of (1) overall cohesion, (2) topology, (3) community structure, and (4) centrality measures to identify influential researchers. We compute these measures using Python's $NetworkX$, $NetworKit$, visualize the networks with R's $ggraph$, and modify a subset of measures based on our operationalization. Cohesion measures include the density, clustering coefficient, average path length, and size of the largest component, which contain details on the overall connectedness of the network, as well as how efficient the network is in facilitating collaborations between researchers from different disciplines, institutions, and countries. 

We also determine whether a co-authorship network follows a power-law degree distribution, indicating a hubs-and-spokes structure where a few hubs accumulate most of the connections. In the co-authorship context, this means that certain key researchers act as central hubs, coordinating the majority of collaborations across the network. To do this, we compute the $\alpha$ goodness-of-fit value, which indicates if a simulated network with the same number of nodes and edges as the co-authorship network exhibits a similar power-law distribution. Generally, a $\alpha$ value between 2 and 3 suggests that a power-law distribution is a good fit \cite{newman2005power}.

To measure structural similarity between the LLM group, ML group, and non-LLM/ML group networks, we conducted comparisons at the node, edge, and network levels. This approach has been validated in prior studies \cite{dinh2024plan,dinh4527922structural}, which found that comparing networks at all three levels because similarities can emerge from local-level interactions and extend to the overall network.

First, we measure node overlap using the normalized pairwise Jaccard index, which is a similarity measure based on the proportion of shared nodes between two networks relative to the total unique nodes in both networks. Second, edge overlap is also calculated using the Jaccard index applied to the adjacency matrices of the networks, so that we can determine the proportion of shared edges relative to the total unique edges. We require that the two networks share at least some common nodes \citep{tantardini2019comparing}, so edge overlap is only calculated when there is a non-zero node overlap.

Third, we compare network-level structural similarity using Netsimile \cite{berlingerio2013network}, a graph embedding method that measures similarity using seven structural features derived from both node- and edge-level properties. Based on the algorithm, we generate a feature vector for each network based on these features: (1) degree, the number of edges connected to each node; (2) clustering coefficient, the proportion of a node's neighbors that are connected to each other; (3) egonet size, the number of edges within a node's neighborhood; (4) egonet density, the ratio of actual to possible edges in each node's ego-network; (5) outward edges, the number of edges connecting the node to others outside its egonet; (6) average neighbor degree, the mean degree of a node's neighbors; and (7) average neighbor clustering coefficient, the mean clustering coefficient of the node's neighbors. We then compare the feature vectors between each pair of networks using Canberra distance, because it is effective at detecting small differences, especially for values near zero, and it normalizes differences to ensure that all features contribute fairly to the overall comparison \cite{berlingerio2013network}. The formula for this measure is below:

\begin{equation}
NetSimileCanberra(G_1,G_2) = \sum_{i=1}^n\frac{\mid{G_{1_i}}- G_{2_i}\mid} {{|G_{1_i}|}+ |G_{2_i}|}
\end{equation}

Here, $G_1$ and $G_2$ are feature vectors for the two networks, with $G_{1_i}$ and $G_{2_i}$ as their $i$th elements. The Canberra distance sums the normalized differences between elements, where smaller values indicate higher network similarity.

\section{Results}\label{sec3}
In \autoref{collaboration}, we analyze the diversity of collaboration based on the institutions and departments affiliated with authors in various fields. In \autoref{network}, we examine the author network to gain insight into how various institutions or departments collaborate, focusing particularly on comparing the collaboration patterns between LLM-related studies and those of control groups. Note that a paper's ``field`` categorization is provided by OpenAlex (see \autoref{openalex} for more details). OpenAlex uses an LLM, specifically a BERT model, to generate a score distribution for identified topics using a paper's title, abstract, and citations \cite{priem2022openalex}. The model provides up to three topics with the highest scores, which we map to fields according to the ASJC structure \cite{priem2022openalex}.

\begin{figure}[htbp]
  \centering
  \includegraphics[width=1\textwidth]{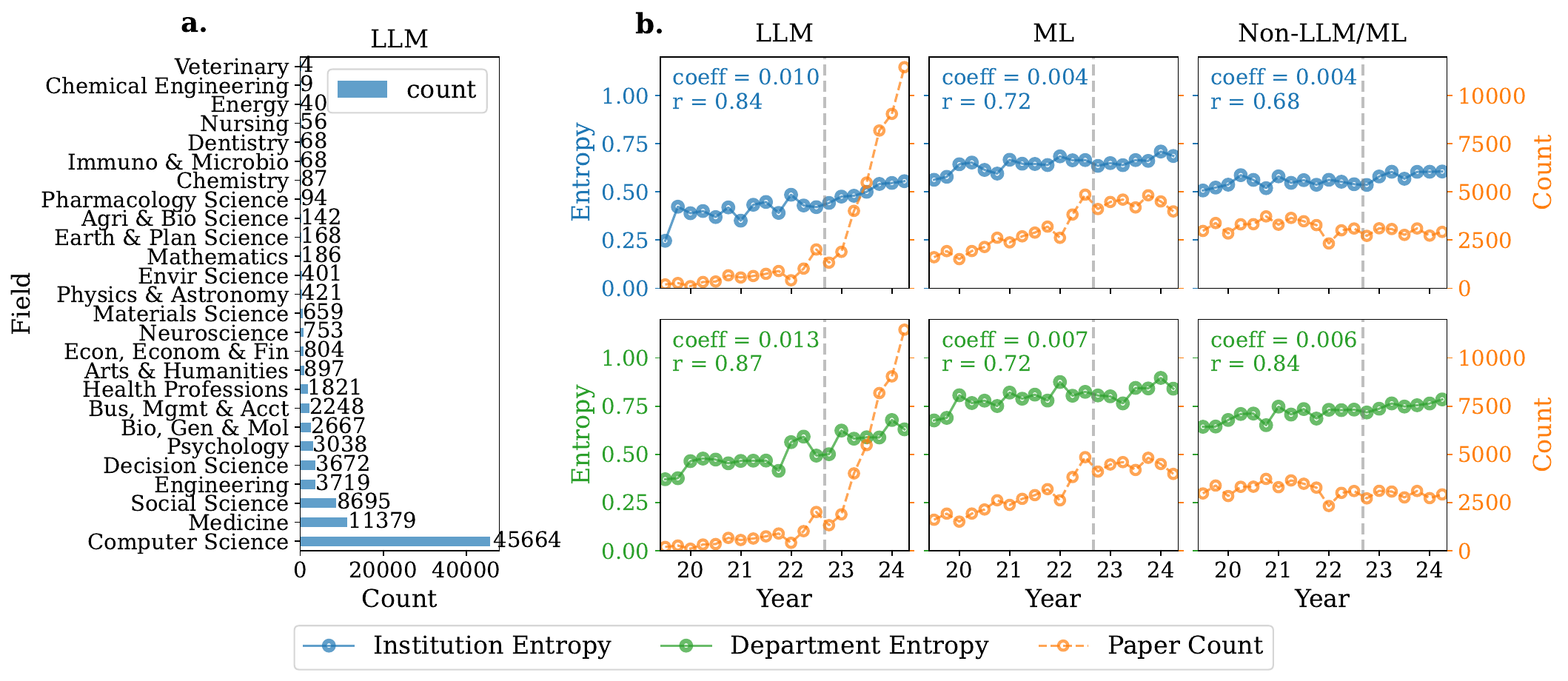}
  \caption{(a) The distribution of papers in the collection by field. (b) The temporal pattern of entropy and paper count based on authors' affiliated institution (top) and department (bottom) information, respectively. In both subplots, the x-axis denotes the date by year. The primary y-axis denotes the Shannon Entropy calculated using \autoref{eq2}, while the secondary y-axis indicates the count of papers. Each point represents the averaged entropy based on papers published in one quarter. The dotted gray line indicates when OpenAI released ChatGPT.}
  \label{fig:distribution}
\end{figure}

We sort the collected 59,293 papers according to the count of papers in order, as presented in \autoref{fig:distribution}(a). For the subsequent analysis, we focus on the top 12 fields with the most publications in the topics of LLMs: (1) \textit{Computer Science}, (2) \textit{Medicine}, (3) \textit{Social Science}, (4) \textit{Engineering}, (5) \textit{Decision Science}, (6) \textit{Psychology}, (7) \textit{Biochemistry, Genetics \& Molecular Biology}, (8) \textit{Business, Management \& Accounting}, (9) \textit{Health Professions}, (10) \textit{Arts \& Humanities}, (11) \textit{Economics, Econometrics \& Finance}, and (12) \textit{Neuroscience}. We then filter for papers with complete authors' affiliation information, resulting in 30,075 papers with complete institution information and 22,453 papers with complete department information. Building on this approach, the ML group and the Non-LLM/ML group include 56,658 and 48,795 papers with institution information, respectively, and 47,002 and 40,579 papers with department information.

Several noteworthy observations emerge from the analysis. First, the Non-LLM/ML group demonstrates consistency in the number of papers, while the ML group exhibits a modest upward trajectory. In contrast, the LLM group shows a pronounced increase in the number of papers, especially after the release of ChatGPT (the dotted gray line in \autoref{fig:distribution}(b)). Second, when examining the coefficient and r-value of the entropy, the LLM group reveals a steeper regression line. Specifically, the entropy increases based on authors' affiliated institution (\(coefficient = 0.010\)) and department (\(coefficient = 0.013\)), which are significant compared to the two control groups.

\subsection{Analysis and comparison of collaboration diversity}
\label{collaboration}

\begin{figure}[htbp]
  \centering
  \includegraphics[width=0.95\textwidth]{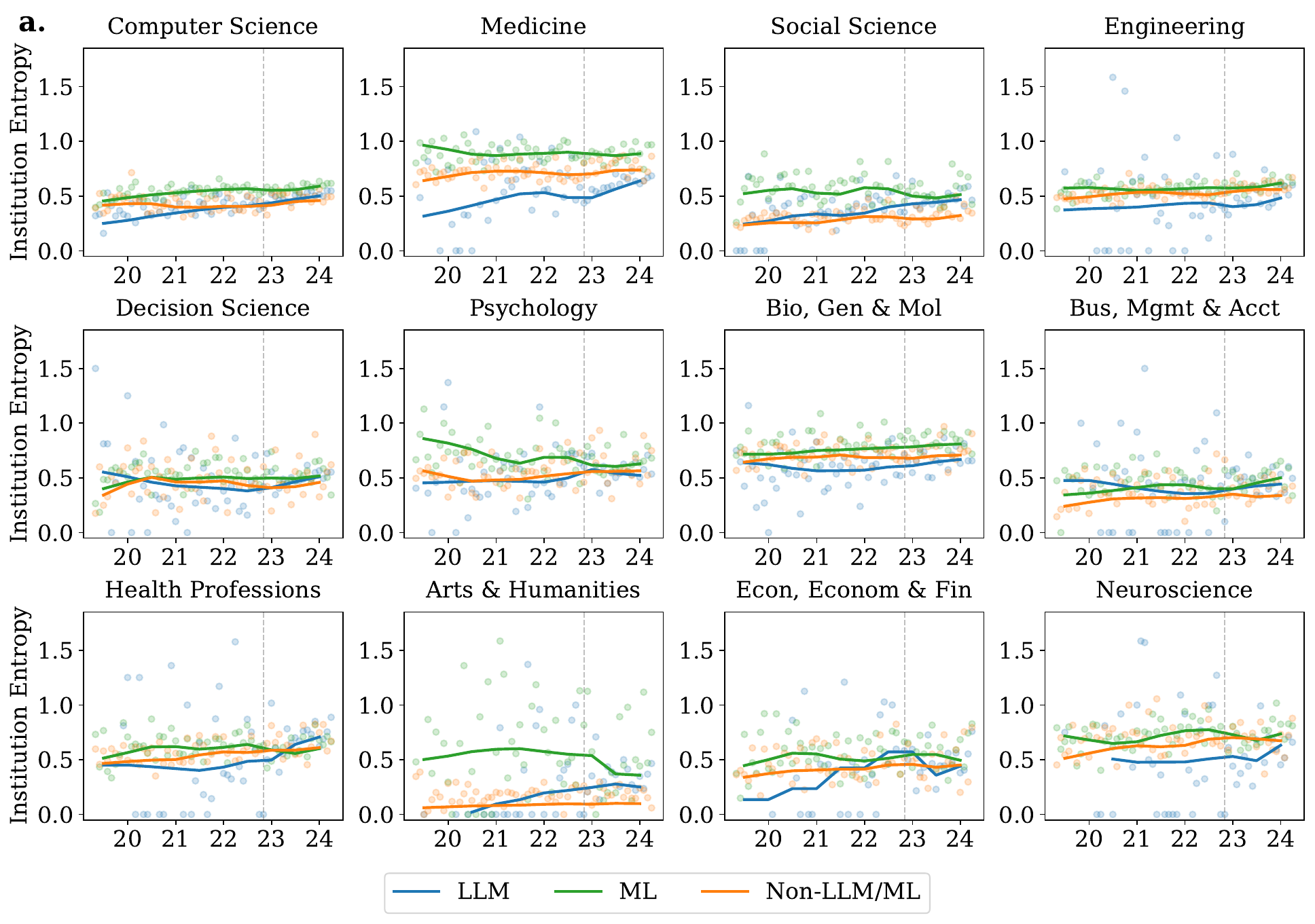}
  \includegraphics[width=0.95\textwidth]{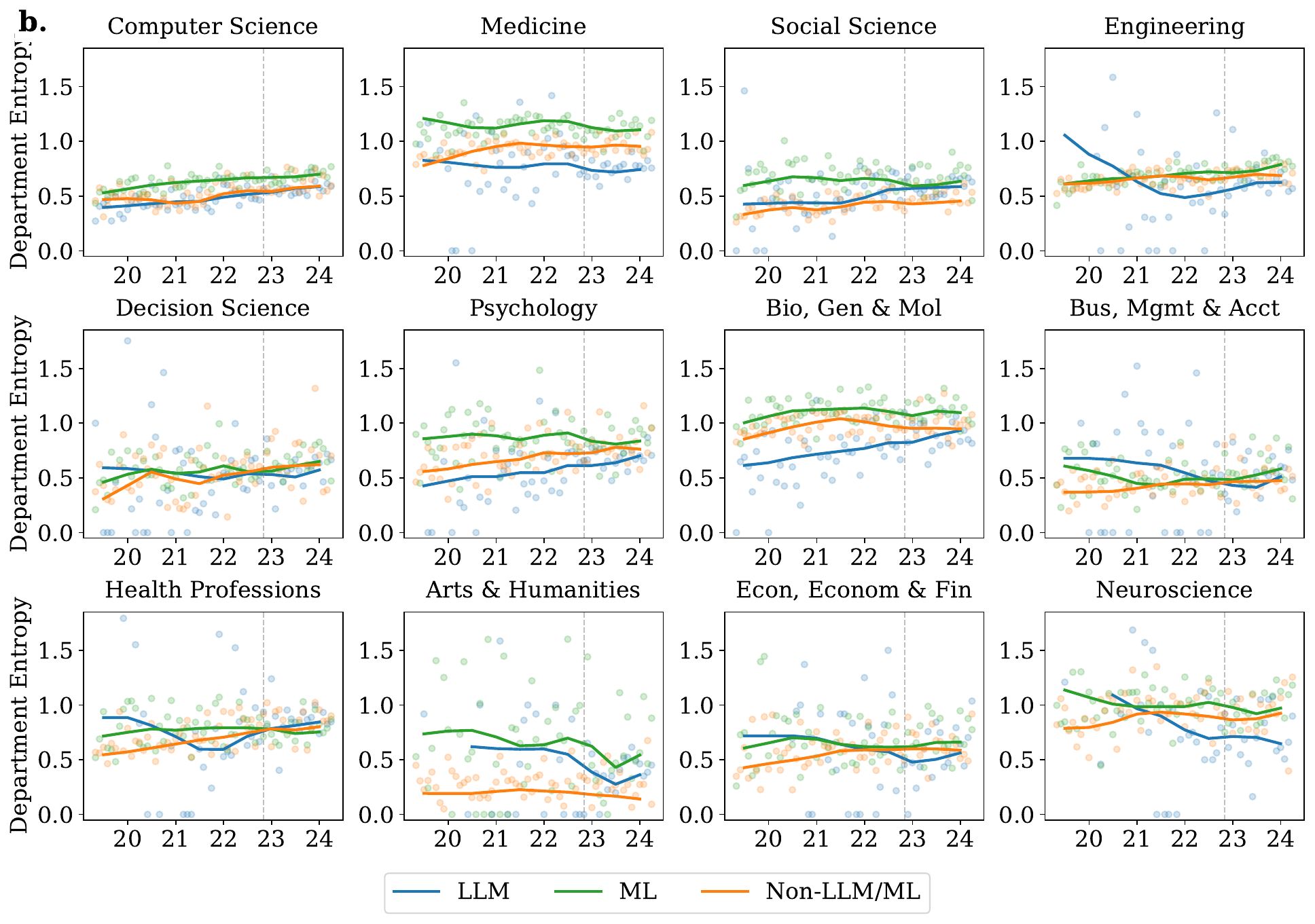}
  \caption{
    Collaboration diversity over time based on (a) authors' institutions and (b) authors' departments. Panel (a) is based on 30,075 papers with complete institution information, while Panel (b) is based on 22,453 papers with complete department information. In both plots, the x-axis denotes the date by year, while the y-axis denotes the Shannon Entropy calculated using \autoref{eq2}. Each point represents the monthly averaged entropy based on papers published in one quarter. The line represents the trend using a non-parametric regression technique called locally weighted scatterplot smoothing (LOWESS). Each point in the panel represents the averaged entropy based on papers published in one year. The dotted gray line implies the date when ChatGPT was released by OpenAI.
  }
  \label{fig:temporal}
\end{figure}

\autoref{fig:temporal}(a) and \autoref{fig:temporal}(b) show the collaboration diversity measured by Shannon Entropy for authors' institutions and departments across three paper groups, respectively. Our analysis in this section addresses the following three key aspects.

\textbf{1. What do the overall collaboration patterns in the LLM group reveal?} Regarding the entropy analysis based on institutions and departments from the LLM group, the Shannon Entropy trends in \autoref{fig:temporal}(a) and \autoref{fig:temporal}(b) (blue line) closely align for each respective field. Most fields show a minor increase in the averaged entropy in 2024. In particular, fields like \textit{Computer Science}, \textit{Medicine}, \textit{Social Science}, and \textit{Biochemistry, Genetics \& Molecular Biology} have shown consistent increases in institutional collaboration. Comparing different fields also reveals distinct patterns. Health-related fields, such as \textit{Medicine}, \textit{Biochemistry, Genetics \& Molecular Biology}, \textit{Health Professions}, and \textit{Neuroscience}, demonstrate higher values in both institution and department entropy levels. This indicates more diverse collaboration patterns across these domains. \textit{Computer Science} exhibits lower overall entropy, particularly in the early stage of LLM studies (e.g., BERT models), suggesting that LLM research initially remained concentrated within the computer science community. Additionally, \textit{Arts \& Humanities} displays notably lower institutional entropy compared to other disciplines.

However, the patterns of collaboration vary across disciplines and even between institutional and departmental entropy within the same field. For instance, in \textit{Engineering}, institutional entropy remains relatively stable while departmental entropy shows high levels in early periods, suggesting that BERT-related research involved cross-departmental collaboration within the same institution. \textit{Medicine} demonstrates a distinct pattern where institutional entropy shows an upward trend, while departmental entropy remains relatively constant. In contrast, \textit{Psychology} and \textit{Biochemistry, Genetics \& Molecular Biology} exhibit reversed patterns compared to \textit{Medicine}.


\textbf{2. Do these collaboration patterns exhibit a general trend comparable to the control paper groups, or are they unique to the LLM paper group?} We compare the entropy for each field across the three groups. Our analysis reveals several consistent patterns. The \textit{ML} group consistently displays higher entropy levels than the other two paper groups; this is particularly pronounced in fields such as \textit{Medicine}, \textit{Computer Science}, \textit{Psychology}, \textit{Social Science}, and \textit{Biochemistry, Genetics \& Molecular Biology}. This likely reflects the well-established nature of research collaborations across disciplines in ML studies. However, when examining the slopes of the smoothing lines, the \textit{LLM} group exhibits a more pronounced increasing trend. This is especially evident in \textit{Computer Science}, \textit{Medicine}, and \textit{Social Science}, where the trend accelerates notably following ChatGPT's release. Additionally, institution entropy remains relatively stable for both the \textit{ML} and \textit{Non-LLM/ML} groups, as compared to the LLM paper group. 

Despite common trends, collaboration patterns vary significantly across disciplines. In \textit{Computer Science} and \textit{Social Science}, the \textit{LLM} paper group demonstrates a more pronounced increasing trend in collaboration, whereas this pattern is less evident in \textit{Engineering} and \textit{Decision Science}. Health-related fields also exhibit distinct patterns: \textit{Medicine} shows a marked increase in institutional collaboration within the \textit{LLM} group, while department entropy remains similar across all three paper groups. In contrast, \textit{Psychology} and \textit{Biochemistry, Genetics \& Molecular Biology} display clearer upward trends in department entropy for the \textit{LLM} group compared to their institution entropy patterns.

\textbf{3. How did the release of ChatGPT influence collaboration dynamics?} Our next question explores the impact of ChatGPT's release. As presented in \autoref{fig:temporal}, the variance of entropy is much wider before ChatGPT's release, possibly due to fewer papers on the topic of LLMs and more diverse collaborations. The averaged entropy is high for some fields like \textit{Engineering}. One possible explanation for this pattern is that researchers from other fields sought partnerships with \textit{Computer Science} experts for such research topics. In addition, the observed changes in entropy suggest dynamics across different fields. Papers in the field of \textit{Computer Science} display a stable increasing trend. \textit{Medicine}, \textit{Social Science}, \textit{Health Professions}, and \textit{Decision Science} also exhibit a notable increase in institution entropy. Other fields display either minor increases or negligible changes. These findings potentially imply the varying impacts of ChatGPT's introduction on interdisciplinary collaborations across different academic fields.

We conduct a Wilcoxon rank-sum test to compare the entropy before and after the release of ChatGPT, with detailed results presented in \autoref{statistics} (see \autoref{fig:test_temporal}). Our analysis reveals statistically significant changes in entropy across several fields. \textit{Computer Science}, \textit{Social Science}, \textit{Psychology}, and \textit{Biochemistry, Genetics \& Molecular Biology} show a statistically significant increase in both institution and department entropies, while \textit{Medicine} and \textit{Neuroscience} display a slight, but not statistically significant, decrease. The other fields do not show any significant change in entropy. In particular, we observe that the mean of the entropy distribution for most fields remains at 0, with the first quartile consistently at 0. This suggests that many researchers primarily collaborate with colleagues from their own institutions or departments on LLM-related topics.

We conduct additional Wilcoxon rank-sum tests to compare entropy across different fields, with results presented as heatmaps in \autoref{fig:test_fields}. Several consistent patterns emerge when examining both institution and department results before and after ChatGPT's release. Certain fields like \textit{Computer Science}, \textit{Medicine}, and \textit{Biochemistry, Genetics \& Molecular Biology} exhibit more significant differences compared to other fields, as evidenced by darker color cells in the heatmaps. Following ChatGPT's release, there is also an increase in significant differences across fields, illustrated by a higher prevalence of darker orange cells in the right heatmap in \autoref{fig:test_fields}, compared to their counterparts in the left heatmap. Another interesting observation is that \textit{Computer Science} consistently demonstrates significantly lower entropy than \textit{Medicine}, \textit{Biochemistry, Genetics \& Molecular Biology}, \textit{Health Professions}, and \textit{Neuroscience}. 

\begin{figure}[h!]
  \centering
  \includegraphics[width=0.7\textwidth]{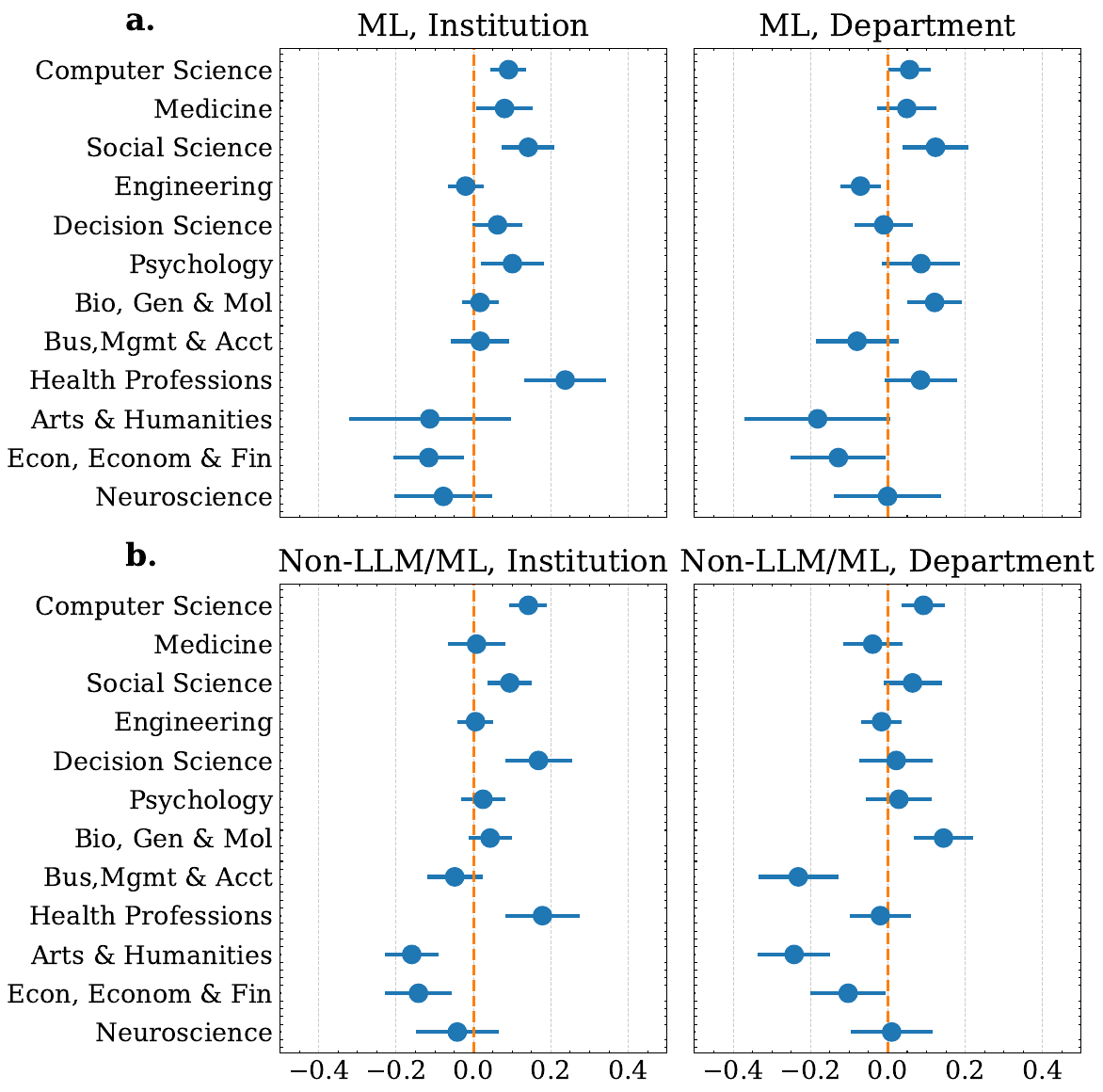}
  \caption{The impact of ChatGPT on the entropy of LLM group as compared to (a) ML and (b) Non-LLM/ML paper groups. Coefficients and Std. Error of DiD ($\beta_3$) by field. When the confidence interval does NOT cross zero (the dashed line), the effect is considered statistically significant, implying that we can be 95\% confident that ChatGPT has a real impact that isn't due to random chance.}
  \label{fig:bsts}
\end{figure}

We further perform difference-in-difference (DiD) analysis to examine the impact of ChatGPT. \autoref{fig:bsts} displays results of $\beta_3$ for the top 12 fields. As shown, the release of ChatGPT consistently leads to a statistically significant increase in entropy in fields such as \textit{Computer Science} and \textit{Social Science}, and a statistically significant decrease in entropy in fields such as \textit{Economics, Econometrics, and Finance}. It also leads to a statistically significant increase in entropy in \textit{Health Professions} and \textit{Decision Sciences} based on institutions, and a statistically significant increase in entropy in \textit{Biochemistry, Genetics, and Molecular Biology} based on departments. Other fields, however, do not show consistent and significant causal impacts from ChatGPT. The strongest positive impacts are mainly observed in health-related fields, such as \textit{Health Professions} and \textit{Biochemistry, Genetics, and Molecular Biology}.

\subsection{Analysis and comparison of collaboration networks}
\label{network}

\begin{figure}[h!]
    \centering
    \includegraphics[width=1\textwidth]{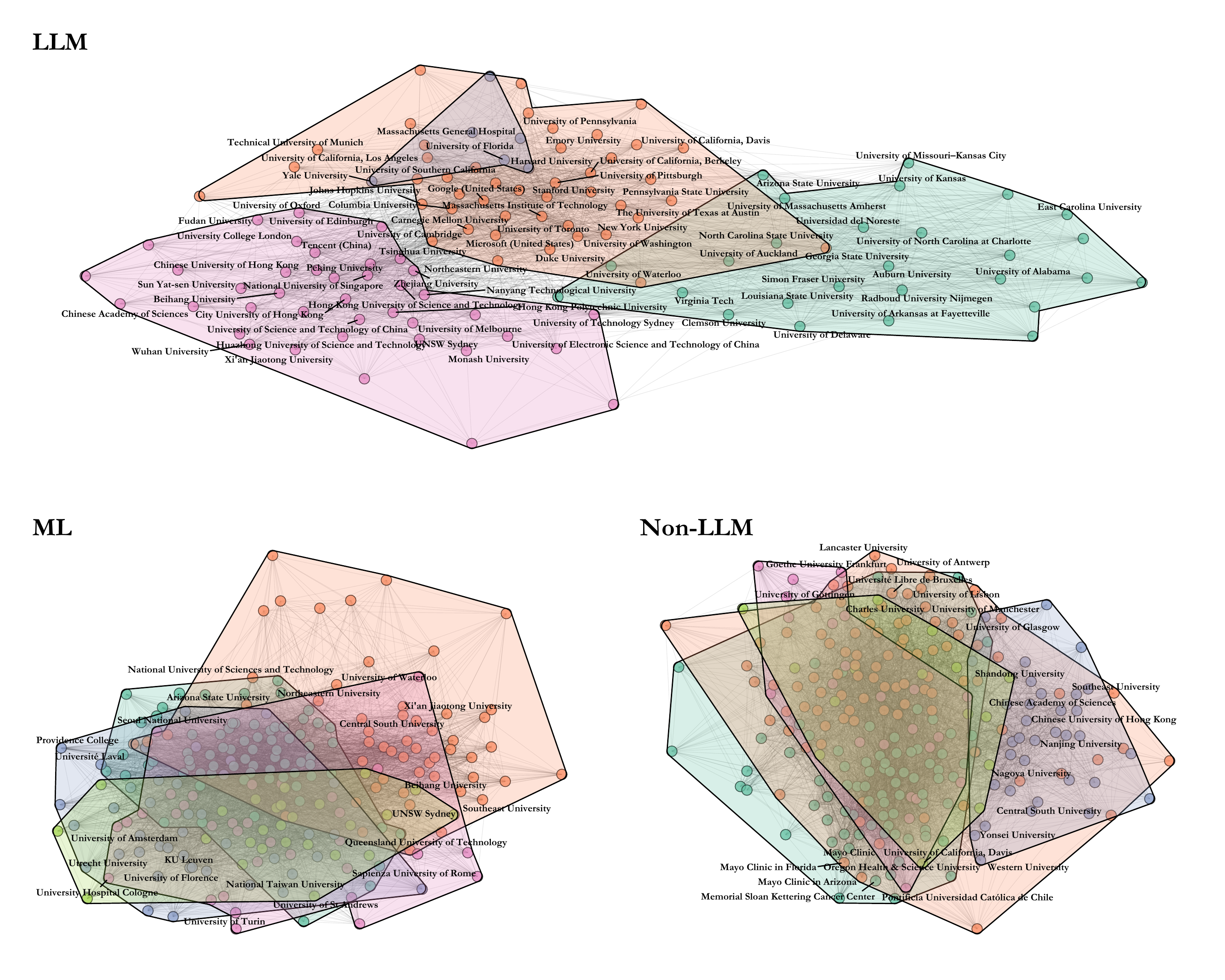}
    \caption{Co-authorship networks based on authors' institutional affiliations. Each node represents a researcher's institution, and each edge represents a co-authorship between pairs of researchers from the respective institutions or departments. Node colors represent the clusters to which the nodes belong, determined based on the Louvain modularity. Node labels represent the top 20\% of nodes ranked by degree centrality. Edge thickness represents the frequency of co-authorship between the connected researchers.}
  \label{fig:networks_institutions}
\end{figure}

\begin{figure}[h!]
    \centering
    \includegraphics[width=1\textwidth]{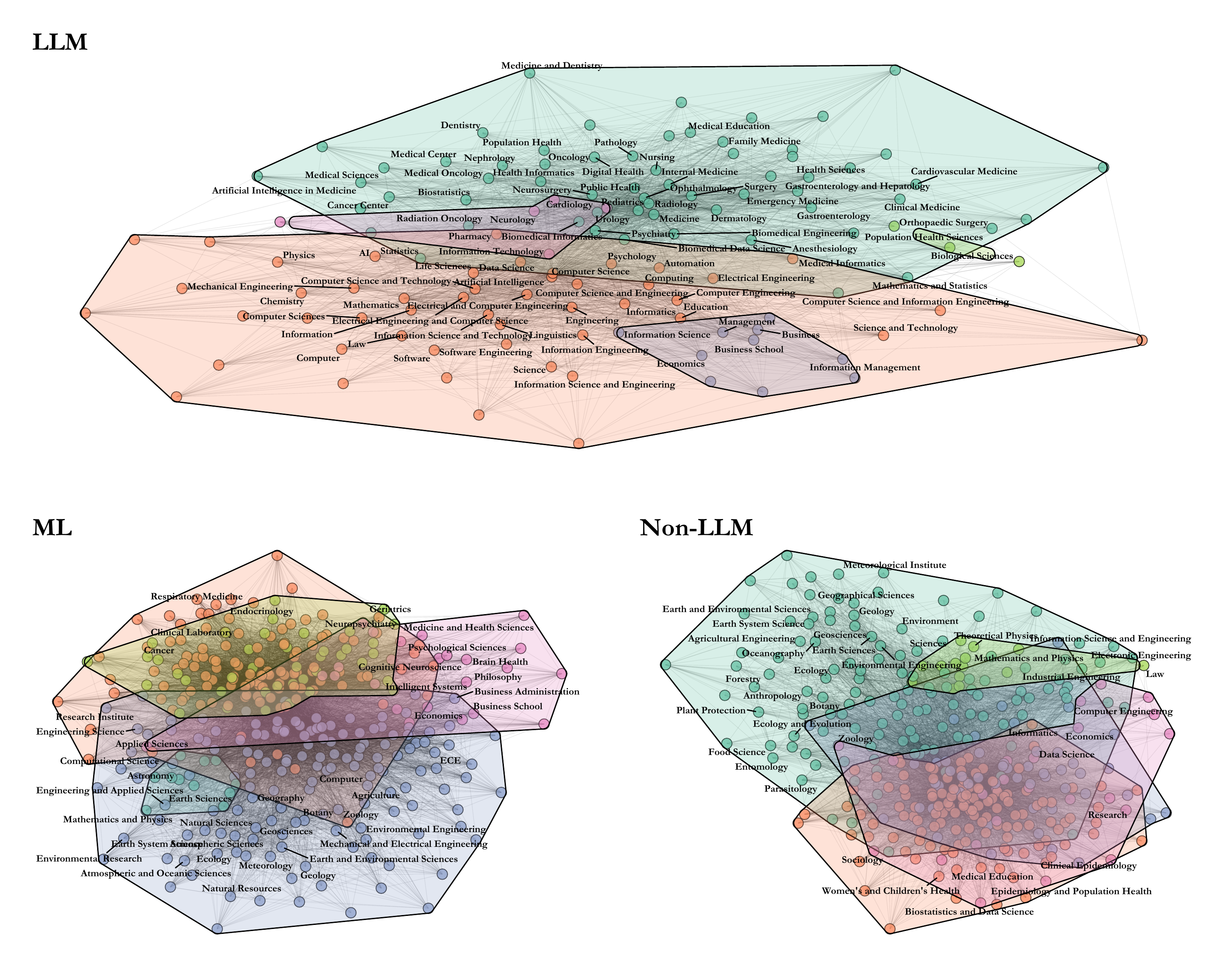}
    \caption{Co-authorship networks based on authors' department affiliations. Each node represents a researcher's department, and each edge represents a co-authorship between pairs of researchers from the respective institutions or departments. Node colors represent the clusters to which the nodes belong, determined based on the Louvain modularity. Node labels represent the top 20\% of nodes ranked by degree centrality. Edge thickness represents the frequency of co-authorship between the connected researchers.}
  \label{fig:networks_departments}
\end{figure}

We analyze the co-authorship networks regarding structural cohesion, topology, community structures, and influential researchers, focusing on dimensions of centrality compared to the control groups. The network metrics appear in \autoref{tab:metrics}, and the network visuals appear in \autoref{fig:networks_institutions}-\autoref{fig:networks_departments}. Our analysis in this section aims to explore the folllowing three questions in terms of collaboration structure. 

\textbf{1. How do the co-authorship networks for LLM papers compare to those in ML and non-LLM/ML research?} Based on \autoref{tab:metrics}, both the institution-based and department-based networks show low density (0.001) and high clustering coefficients (0.60 and 0.79, respectively), but existing ones form tight-knit clusters. The largest components make up a significant part of all network edges (78\% for the institutions network and 76\% for the departmental affiliation network). This suggests that most researchers belong to research communities that can reach one another. The control groups' networks resemble the LLM network, exhibiting low density (range: 0.0003-0.001) and high clustering coefficients (range: 0.59-0.81). 

The relatively high average shortest path length (14.37 for institution-based, and 18.57 for department-based networks) suggests extensive reach within the networks, indicating that while direct connections between researchers might be limited, there are alternative paths that connect them. The LLM-based network, similar to the control group's networks, also exhibits a stronger fit towards a power-law distribution ($\alpha$ range: 2.24-3.75), suggesting that central departments, namely \textit{Computer Science}, and \textit{Medicine}, are hubs that facilitate numerous collaborations across all networks, regardless of whether the collaborations are on topics that are LLM-related, ML-related, or non-LLM-related. \autoref{tab:top5_centrality_departments} reveals that the top central nodes across domains primarily consist of computer science, medicine, and their related subdisciplines. 

\textbf{2. How do the top institutions and departments in LLM papers differ from those in ML and non-LLM/ML research?}
Degree centrality results highlight the prominence of \textit{Computer Science}, \textit{Medicine}, and other \textit{Computer Science}-related departments in LLM research. Furthermore, the high betweenness centrality of these same departments highlights their roles in facilitating collaboration between departments that would have otherwise been unconnected. As shown in \autoref{fig:networks_departments}, the clusters of departments between co-authors not only demonstrate the dominance of \textit{Computer Science} and \textit{Medicine} and their related disciplines (the \textcolor{cscluster}{orange} cluster and the \textcolor{medicinecluster}{green} cluster for \textit{Computer Science} and \textit{Medicine}-related disciplines, respectively) in the network, but also shows that these two disciplines connect fields with little to no overlap. For instance, \textit{Medicine} is in the shortest path between \textit{Engineering} and \textit{Social Science} in 196 instances. Similarly, \textit{Computer Science} is in the shortest path between \textit{Medicine} and \textit{Psychology and Neuroscience} for 384 instances.

The betweenness centrality results reveal notable differences, with international institutions, such as \textit{Instituto de Ciencias de La Construcción Eduardo Torroja} and \textit{Heilongjiang Academy of Sciences}, serving as key bridges in the LLM network. These institutions connect otherwise disconnected research communities, particularly linking major hubs like \textit{Chinese Academy of Sciences}, \textit{Tsinghua University}, and \textit{University of Amsterdam} with smaller or regional institutions. This contrasts with the ML and non-LLM/ML networks, where well-established institutions, primarily from the U.S. and U.K., dominate. At the department level, the LLM network highlights more specialized areas, such as \textit{Dynamics, Logics and Inference for Biological Systems} and \textit{Data and Knowledge Management}, while traditional departments like \textit{Computer Science} and \textit{Medicine} continue to hold central positions in the ML and non-LLM/ML networks, reflecting their long-standing influence between communities of research. 


Closeness centrality results show how researchers from departments and institutions are, on average, reachable to other researchers. In the LLM network, prominent colleges across U.S. (\textit{Harvard University}), the U.K. (\textit{UCL}), and China (\textit{Peking University}) are on average, closest to all other institutions in the network. On the other hand, the ML and non-LLM/ML networks are more fragmented, with smaller or regional institutions such as \textit{Trilogi University} and \textit{Universitas Bina Darma} with the highest closeness scores. At the department level, \textit{Computer Science} and \textit{Medicine} are the most central in the LLM network, while in the ML and non-LLM/ML networks, Computer Science-related departments dominate, with no Medicine-related departments appearing as central in terms of closeness centrality.


Eigenvector centrality results show more consistency with degree centrality and betweenness centrality results than with closeness centrality, highlighting researchers who are not only well-connected but also connected to other influential researchers. Institutions with the highest eigenvector centrality scores are also those with high degree and betweenness centrality, namely \textit{Stanford University} and \textit{Tsinghua University}.


\begin{table}[h!]
\centering
\caption{Comparison of network metrics based on authors' institutions and departments}
\begin{tabular}{cp{1.1cm}p{1.1cm}p{2.2cm}p{1.1cm}p{1.1cm}p{2.2cm}}
    \toprule
    \textbf{Metric} & \multicolumn{3}{c}{\textbf{Institutions}} & \multicolumn{3}{c}{\textbf{Departments}} \\ \hline
     & \textbf{LLM} & \textbf{ML} & \textbf{Non-LLM/ML} & \textbf{LLM} & \textbf{ML} & \textbf{Non-LLM/ML} \\ 
    \midrule
    No. Nodes & 9320 & 16391 & 19010 & 10131 & 24707 & 26492 \\ \hline
    No. Edges & 40971 & 119960 & 104649 & 33314 & 121854 & 94901 \\ \hline
    Density & 0.0009 & 0.0009 & 0.0006 & 0.0006 & 0.0004 & 0.0003 \\ \hline
    Avg. Shortest Path & 14.37 & 13.4 & 15.36 & 18.57 & 16.47 & 22.79 \\ \hline
    Clustering Coefficient & 0.6 & 0.59 & 0.6  & 0.79  & 0.8 & 0.81 \\ \hline
    No. Components & 1729 & 1481 & 3113 & 1963 & 2786 & 4686 \\ \hline
    Largest Component & 7266 & 14699 & 15426 & 7679 & 21261 & 20649 \\ \hline
    Power-law Exponent & 3.75 & 2.29 & 2.54 & 2.29 & 2.24 & 2.31 \\ \hline
    No. Communities (Louvain) & 1774 & 1508  & 3150 & 2010 & 2833 & 4729 \\ \hline
    No. Communities (CNM) & 1856 & 1707 & 3308 & 2091 & 3126 & 5066 \\ 
    \bottomrule
\end{tabular}
\label{tab:metrics}
\end{table}

\textbf{3. How similar (or different) are LLM networks to ML and non-LLM/ML networks in node, edge, and structural overlap?
}
At the institutional level, the LLM and ML networks share notable node overlaps, such as \textit{Stanford University}, \textit{Tsinghua University}, and \textit{University College London}, which appear as central nodes in both networks. However, the edge Jaccard similarity remains low (0.025), showing that while these institutions are present in both networks, their patterns of collaboration differ significantly. In contrast, the LLM network shares fewer nodes with the non-LLM/ML network, as seen in the lower node Jaccard similarity (0.293), with some overlap in institutions like the \textit{Chinese Academy of Sciences} and \textit{Peking University}. At the overall network level, the Netsimile distance indicates that the LLM network is more similar to the non-LLM/ML network (3.203) compared to the ML network (6.275), suggesting that LLM research draws from a broader institutional base rather than mirroring the more concentrated ML collaborations.

At the department level, node and edge Jaccard values are consistently lower than those at the institutional level, with LLM to ML showing slightly higher overlap (0.141 for nodes, 0.03 for edges) compared to LLM to non-LLM/ML (0.113 for nodes, 0.02 for edges). This indicates that while Computer Science is central across all networks, the LLM network includes more specialized departments like \textit{Artificial Intelligence}, \textit{Data Science}, and \textit{Information Science}, which do not appear prominently in the ML and non-LLM/ML networks. In contrast, the non-LLM/ML network contains nodes representing traditional fields such as \textit{Physics} and \textit{Biological Sciences}, while the ML network is concentrated around \textit{Computer Science}, \textit{Engineering}, and \textit{Electrical Engineering}. The Netsimile distance further highlights these differences, showing greater structural similarity between LLM and non-LLM/ML networks (2.264) than between LLM and ML networks (6.35). This suggests that ML-driven collaborations are more tightly clustered around established technical fields, whereas the LLM network, like the non-LLM/ML network, it contains clusters from a broader range of domains, spanning computational sciences to environmental sciences and business.

\begin{table}[h!]
    \centering
    \caption{Comparison of node Jaccard similarity, edge Jaccard similarity, and Netsimile distance}
    \begin{tabular}{lccc} 
        \toprule
        \textbf{Comparison} & \textbf{Node Jaccard Similarity} & \textbf{Edge Jaccard Similarity} & \textbf{Netsimile Distance} \\ 
        \midrule
        \multicolumn{4}{c}{\textbf{Institutions}} \\ \hline
            LLM to ML  & 0.332 & 0.025 & 6.275 \\ \hline
            LLM to non-LLM/ML & 0.293 & 0.019 & 3.203 \\ \hline
        \multicolumn{4}{c}{\textbf{Departments}} \\ \hline
            LLM to ML  & 0.141 & 0.03 & 6.35 \\ \hline
            LLM to non-LLM/ML  & 0.113 & 0.02 & 2.264 \\ 
        \bottomrule
    \end{tabular}
    \label{tab:network_comparisons}
\end{table}

\section{Discussion}
\label{sec4}

\subsection{Key findings and implications}

Collaboration diversity results, as well as co-authorship structures, reveal that LLMs have notable and complex effects on interdisciplinary and cross-institutional collaborations. Integrating LLMs into research areas has the potential to shift collaboration patterns. Below, we summarize several key findings and implications.

\textbf{First, scientific collaboration generally increases in frequency, but the specific patterns vary by discipline.} Since the advent of ChatGPT in 2022, the number of publications on LLM-related topics has increased significantly. This surge extends beyond \textit{Computer Science} to fields such as \textit{Medicine}, \textit{Social Science}, and \textit{Engineering}. Additionally, the overall entropy, which measures collaboration diversity, has increased. This observation implies that collaboration diversity has broadened across fields that employ LLMs in research. 

However, the entropy pattern on specific fields varies. \textit{Computer Science} sees a stable increase in entropy, suggesting that computer scientists continue to seek collaboration with researchers from other disciplines. This trend aligns with the observation that researchers explore topics such as AI for social good and AI for science \cite{tomavsev2020ai}. \textit{Social Science} also exhibits a significant rise in both institution entropy and department entropy following the release of ChatGPT. This likely reflects the growing application of computational methods and AI tools among social scientists to conduct quantitative analyses \cite{ziems2024can}. Health-related fields, including \textit{Medicine}, \textit{Neuroscience}, \textit{Health Professions}, and \textit{Biochemistry, Genetics, \& Molecular Biology}, exhibit higher entropy compared to \textit{Computer Science}, \textit{Social Sciences}, and \textit{Engineering}. Among these health-related fields, \textit{Psychology}, \textit{Biochemistry, Genetics, \& Molecular Biology}, and \textit{Health Professions} have experienced particularly notable increases in institution entropy. Our observation aligns with prior research showing that researchers in health are enthusiastic about using LLMs to facilitate documentation of patient reports, improve diagnosis accuracy, and assist in clinical care \cite{peng2023study}. It is worth noting that \textit{Computer Science} exhibits lower entropy overall compared to health-related fields. One possible explanation is that health-related articles often have longer author lists. Additionally, in the early stages of LLM research, researchers from various disciplines sought collaborations with Computer Science researchers. This focus aligned with the fact that LLMs were initially concentrated within the domain of Computer Science, as it was a highly specialized topic during that period.



\textbf{Second, scientific collaboration within the LLM paper group indicates potential for increased interdisciplinary collaboration.} This trend is particularly evident in fields such as \textit{Computer Science}, \textit{Medicine}, \textit{Social Science}, and \textit{Health Professions}. In general, AI-related studies exhibit higher levels of collaboration, with ML paper groups consistently positioned above Non-LLM/ML paper groups. However, the entropy of the LLM paper group starts at a much lower level, indicating a higher entry barrier in earlier stages. After the release of ChatGPT, collaboration across most fields demonstrates an upward trend, and the gap between the LLM paper group and the ML paper group narrows. 

A plausible explanation is that \textit{Computer Science} researchers increasingly seek partnerships with domain experts or are sought by their colleagues focusing on interdisciplinary areas such as AI for scientific applications or AI for social good. In other fields, the advancements in LLMs may have lowered entry barriers or introduced new opportunities in domain-specific areas. For instance, domain-specific studies in pathology \cite{lu2024multimodal} and clinical diagnosis \cite{peng2023study}, which evaluate whether LLM-predicted outputs correlate with improved patient outcomes, likely contribute to the growth of interdisciplinary collaboration.

\textbf{Third, the advancement of LLMs has a significant impact on scientific collaboration.} 
Our analysis indicates that these advancements like the release of ChatGPT have potentially lowered the entry barriers and broadened the applications of LLMs across various fields. Consequently, most disciplines have shown a substantial increase in the number of publications. Furthermore, many fields, including \textit{Computer Science} and \textit{Social Science}, have witnessed a consistent increase in the diversity of collaborations.

For other fields like \textit{Medicine} and \textit{Neuroscience}, inter-institutional collaboration has grown, but intra-departmental collaboration has not increased similarly. While LLMs have facilitated broader collaborations across institutions, they may also promote specialized collaborations within certain domains where expert knowledge is critical. For example, medical research often involves stringent data privacy requirements, such as HIPAA regulations for electronic health records \cite{jiang2023health}. Unlike earlier AI models, researchers in \textit{Medicine} can now leverage LLMs to address domain-specific challenges more effectively, often without requiring external collaborations from other departments.


\textbf{Fourth, structural analysis of co-authorship connections shows that \textit{Computer Science} and \textit{Medicine} remain the most represented disciplines in the network.} In particular, researchers from these two disciplines have the highest degree and betweenness centrality scores, indicating their roles as both active researchers and facilitators of cross-field collaborations. This finding holds special importance for the field of \textit{Computer Science} because, despite their papers featuring fewer disciplines (indicating less entropy compared to others), they play a crucial role in connecting various disciplines. For instance, \textit{Medicine} bridges \textit{Computer Science} and \textit{Engineering and Information Technology}, \textit{Engineering} and \textit{Social Science}, and \textit{Neurosurgery} and \textit{Intelligent Technology and Engineering}. \textit{Medicine}'s influence is also reflected in the cross-institutional analysis, where medical institutions and associated national laboratories (e.g. \textit{University of Colorado Anschutz Medical Campus}, \textit{European Bioinformatics Institute}, {Jackson Laboratory}) are the most influential in terms of eigenvector centrality, highlighting their central roles in connecting other influential institutions and facilitating extensive cross-institutional collaborations.

\textbf{Last, structural analysis reveals that the advancement of LLMs has the potential to foster interdisciplinary collaboration.} The LLM collaboration network preserves network properties similar to established research fields while showcasing a unique democratization of research participation. International institutions and specialized departments play prominent roles as key bridges within the network, marking a significant shift from the traditional ML research paradigm, which historically centered around leading U.S. and U.K. institutions. The lower Jaccard similarity values between LLM and ML networks suggest that ChatGPT has sparked potential cross-disciplinary collaborations, reaching beyond traditional ML research boundaries. Additionally, the high clustering coefficients, coupled with the network's extensive reach, indicate that researchers still collaborate within their immediate communities. However, ChatGPT's versatility has enabled researchers to forged connections between previously disconnected domains. This shift has fostered a more distributed and interdisciplinary research ecosystem. These findings highlight how ChatGPT has not only democratized access to LLM research but also actively facilitated the growth of interdisciplinary collaborations across diverse fields.

Overall, the study of author collaboration and network analysis provides valuable insights into the dynamics and patterns of interdisciplinary research in LLMs, highlighting how knowledge and expertise are exchanged across various fields. Our analysis reveals that the diversity of academic collaborations increases overall, but varies across disciplines. Fields with methodological expertise and domain generality like \textit{Computer Science} exhibit an increasing collaboration diversity. In contrast, disciplines like \textit{Medicine} exhibit decreased collaboration diversity after the release of ChatGPT. This decline likely results because LLMs lower the demands for external technical and/or methodological expertise. Two dominant fields, Computer Science and Medicine, facilitate extensive cross-institutional and cross-country collaborations. Leading research institutions and associated national labs in the U.S., U.K., and China actively conduct research using LLMs, fostering interdisciplinary partnerships while sharing expertise and resources.

\subsection{Opportunities for future work}

This study presents several avenues for future research, each addressing current limitations and opportunities for expansion. One primary constraint lies in the data quality of OpenAlex, which has been shown to have missing information issues, particularly in author affiliation and abstract details \cite{zhang2024missing}. To mitigate this, our analysis excludes records with incomplete data. Consequently, approximately 50\% of the papers retain complete information on authors' institutions, and about 40\% retain detailed department information. While this approach helps address the data quality issue, it also leads to the exclusion of valuable insights from papers or preprints that are not fully captured by OpenAlex, especially in fields with fewer published papers. Nevertheless, as highlighted in prior research \cite{cespedes2024evaluating}, OpenAlex continues to demonstrate better inclusiveness and coverage of academic papers compared to other databases. Future research could benefit from conducting sensitivity analyses to assess the impact of these missing values on the patterns observed in this study, thereby providing a more robust understanding of the data's reliability and comprehensiveness.

The publication timing of a paper could potentially influence our results. In this study, we use the publication time of a paper as the reference for time. However, for papers without a specific publication date---particularly those from Computer Science conferences, which often only provide the publication year by OpenAlex---we utilize the created\_date field from the OpenAlex dataset as a proxy for the publication time. This approach ensures the inclusion of papers from conference tracks while providing a reasonable approximation for their publication dates. Future research could incorporate the information of papers from Computer Science conference tracks available in their respective databases. However, ensuring consistency during the data fusion process presents an additional challenge that would need to be addressed.

Another area for improvement concerns OpenAlex's categornization of disciplines. While they utilize a BERT-based model to identify topics and subsequently map them to fields based on Scopus's ASJC structure (see \autoref{openalex}), there is a lack of clear statement regarding the training and testing data size, as well as the reported performance metrics for discipline classification. A valuable direction for future work would be to evaluate the accuracy and reliability of this discipline classification system, potentially developing more refined or domain-specific categorization methods for LLM-related research.

Next, future research could focus on continuously collecting papers and expanding the scope to encompass the broader concept of generative AI, such as multimodal LLMs (e.g., GPT-4o). Given that the landscape of LLMs has undergone dramatic changes, with new models and breakthroughs emerging at an unprecedented pace, this work could allow for a more comprehensive longitudinal analysis of trends, innovations, and shifts in academia. Additionally, we could explore the interplay between academic research and industry developments and track how breakthroughs in LLMs influence and are influenced by practical applications. This could provide valuable insights into the trajectory of LLM and generative AI development and help identify emerging subfields, interdisciplinary connections, and potential areas for future breakthroughs.


\backmatter








\section*{Declarations}


\begin{itemize}
\item Conflict of interest/Competing interests. \\ 
The author declares no competing interests.
\item Data availability. \\
The data files can be accessed at: https://doi.org/10.5281/zenodo.13118978
\item Code availability. \\
The code files can be accessed at: https://github.com/Lingyao1219/llm-science
\end{itemize}



\begin{appendices}

\section{Data Preparation}
\label{secA}

\subsection{Data preparation workflow}
\label{workflow}
\begin{figure}[h!]
  \centering
  \includegraphics[width=1\textwidth]{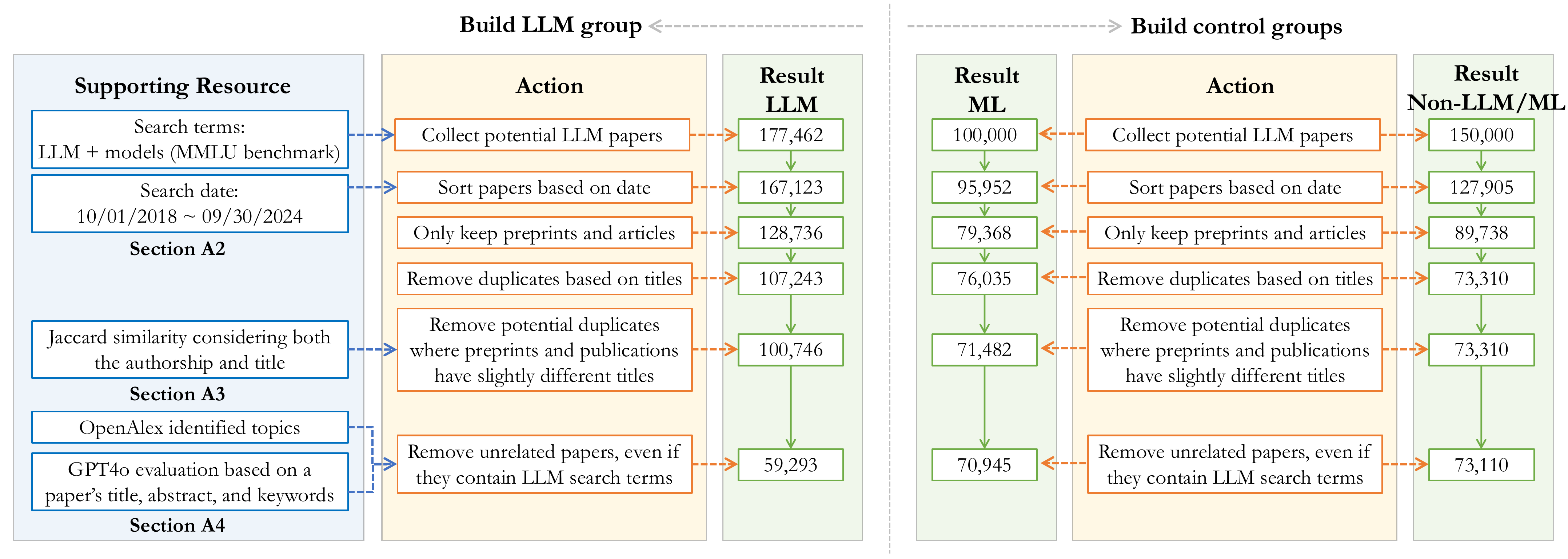}
  \caption{An illustrative workflow for paper collection and cleaning.}
  \label{fig:cleaning}
\end{figure}

\autoref{fig:cleaning} outlines the process for collecting and cleaning the dataset of LLM-relevant papers for our analysis. The process begins with a broad search using general terms related to LLMs and popular models based on the MMLU benchmark (see \autoref{search}) \cite{hendrycks2020measuring}, spanning from October 2018 to September 2024. We apply this search to the title and abstract to avoid excessive noise in the dataset. This initial search yields a collection of 177,462 papers, which then undergo a series of filtering steps to enhance relevance and remove duplicates.

The cleaning process involves several steps, each progressively narrowing down the dataset. First, the collection is limited to preprints and articles, reducing it to 128,736 papers. The next step involves removing duplicates based on titles. Additionally, since a preprint might change its title upon official publication, we examine papers with slight variations between preprints and formal publications using Jaccard similarity (see \autoref{jaccard}). After reviewing the papers with Jaccard similarity, we find that a paper with a similarity score greater than 0.6 is very likely to be duplicated, and we remove those potential duplicates. However, it is still possible for a paper to contain LLM-related keywords but not relevant to LLM studies, such as the keyword ``PaLM'' which often appears in papers discussing ``palm trees'' or ``palm oil.'' To address this issue, we employ the GPT-4o model to support the evaluation of the relevance of a paper (see \autoref{relevance}). This filtering process results in a final set of 59,293 papers, representing a focused and highly relevant collection for our analysis. The distribution of these papers by field is presented in \autoref{fig:distribution}.

\subsection{Paper collection}
\label{search}
\autoref{tab:llm_search_terms} below shows the specific search terms used to collect papers from OpenAlex. The search terms include two general terms, ``large language model" and ``LLM," along with popular open-source models (e.g., BERT, Flan-T5, LLaMA) and closed-source models (e.g., ChatGPT, Claude) based on the MMLU benchmark \cite{hendrycks2020measuring}. It should be noted that our search term like ``GPT'' encompasses all relevant models, including GPT-3, GPT-4, and GPT-4o.

\begin{table}[h!]
    \centering
    \caption{LLM-related search terms}
    \begin{tabular}{p{0.2\textwidth}p{0.7\textwidth}}
        \toprule
        \textbf{Category} & \textbf{Search terms} \\
        \midrule
        General LLM terms & (Large language model) OR LLM \\
        \hline
        Close-sourced & GPT OR ChatGPT OR Claude OR Gemini OR (PaLM 2) OR (davinci-002) OR (davinci-003) OR Chinchilla \\
        \hline
        Open-sourced & BERT OR RoBERTa OR LLaMA OR Mistral OR Mixtral OR Qwen OR DBRX OR Falcon OR BLOOM OR (Flan T5) OR (Flan PaLM) OR LaMDA \\
        \bottomrule
    \end{tabular}
    \label{tab:llm_search_terms}
\end{table}

The control groups follow the same process. For the ``ML'' control group, we randomly select 100,000 papers that containing the keyword ``machine learning'' from the same time periods and apply the processing pipeline. This process reduces the dataset to 71,482 papers. To ensure that the collected ML papers are distinct from those in the LLM group, we remove any overlapping papers that appear in the LLM group. Similarly, for the ``Non-LLM/ML'' papers, we initially collect a random sampling of 150,000 papers. Then, we follow the same cleaning procedure and eliminate overlaps. In other words, each group is mutually exclusive, with no papers belonging to more than one group. As a result, we have 59,293, 70,945, and 73,110 papers for ``LLM,'' ``ML,'', and ``Non-LLM/ML'' groups, respectively.

\subsection{Duplicates removal}
\label{jaccard}

We first remove the duplicated papers based on the title information. This handling reduces the collection of papers to 128,736. Given that a preprint might change its title when officially published, we then use the Jaccard similarity coefficient to identify works with variations in titles but are actually the same. The Jaccard similarity coefficient is a measure of similarity between two sets. It is defined as the size of the intersection divided by the size of the union of the sets, as shown below:

\begin{equation}
J(A, B) = \frac{|A \cap B|}{|A \cup B|}
\end{equation}

In our context, sets \(A\) and \(B\) represent the collections of words from the titles and author names of a preprint and its paired articles, respectively. \(|A \cap B|\) denotes the number of common words between the titles and author names of the preprint and its paired articles, while \(|A \cup B|\) is the total number of unique words in both titles and author names combined. We output all works with \(J(A, B) \geq 0.5\) and find that \(J(A, B) \geq 0.6\) often implies the preprint and the published article are the same paper. Therefore, we remove all those preprints showing a \(J(A, B) \geq 0.6\).

\subsection{Relevance evaluation}
\label{relevance}
This step involves filtering out papers that are not actually discussing LLM-related topics, even if their abstracts contain key terms such as ``Palm.'' We take several factors into consideration for this filtering process. First, we consider papers with identified topics (provided by OpenAlex) including ``natural language processing,'' ``artificial intelligence,'' ``machine learning,'' ``text mining,'' ``deep learning,'' ``transfer learning,'' ``question answering,'' and ``speech recognition'' as LLM-relevant. Then, we use GPT-4o mini to determine their relevance based on the title, abstract, and topics. The prompt is designed as follows:

\begin{lstlisting}
Please identify if the following paper is related to the topic of large 
language models (LLMs) or involves the use of LLMs based on the 
following provided information.
----------
Paper Title: {title}
Abstract (if available): {abstract}
Topics (if available): {topics}
----------
If the abstract is unavailable, please use the title and topics to make 
your determination. Please note that a paper mentioning concepts like 
`neural network', `machine learning', `artificial intelligence', or 
any NLP tasks always suggests a connection to LLMs. For example, a 
paper titled `The improved neural network model in humor detection' 
is very likely to involve LLMs. Respond `Yes' for such papers. Please 
only respond with either `Yes' or `No'. Please do not return other 
output.
\end{lstlisting}

To validate the classification, we manually review 190 randomly selected papers, of which 86 are classified as irrelevant and 104 as relevant based on our manual annotation. We then compare these results with those returned by GPT-4o mini. We use F1-score and accuracy to evaluate performance. Overall, GPT-4o mini's identification achieves an agreement of 0.96 with manual annotation, with F1-scores of 0.96 for both irrelevant and relevant classes.

\subsection{Discipline-related information extraction}
\label{extraction}

As the department information is not directly provided by OpenAlex, this step involves extracting discipline-related information from the authors' affiliation information, which allows us to conduct department analysis in the result section. Again, we use GPT-4o mini and design the following prompt to facilitate the information retrieval. For example, if an author's affiliation information is ``School of Computer Science, Peking University, Beijing, China,'' then this prompt allows us to extract the information ``Computer Science'' as the relevant discipline. It is important to note that we do not merge closely related disciplines, such as ``Computer Science'' and ``Computer Engineering,'' to ensure precise and unbiased extraction.

\begin{lstlisting}
Please extract the discipline-related information from the following 
author's affiliation: 
----------
Author's affiliation: {affiliation}
----------
For example, for 'Department of Biological Science, Joseph Ayo Babalola 
University, Nigeria', return 'Biological Science'. If it is not written 
in English, translate it to English and return. If you cannot identify 
any discipline-related information, return 'None'. 
Please do not return other output or explanation.
\end{lstlisting}

\section{Statistical Analysis}
\label{statistics}

\subsection{Wilcoxon rank-sum test}
\label{wilcoxon}

\begin{figure}[h!]
  \centering
  \includegraphics[width=0.95\textwidth]{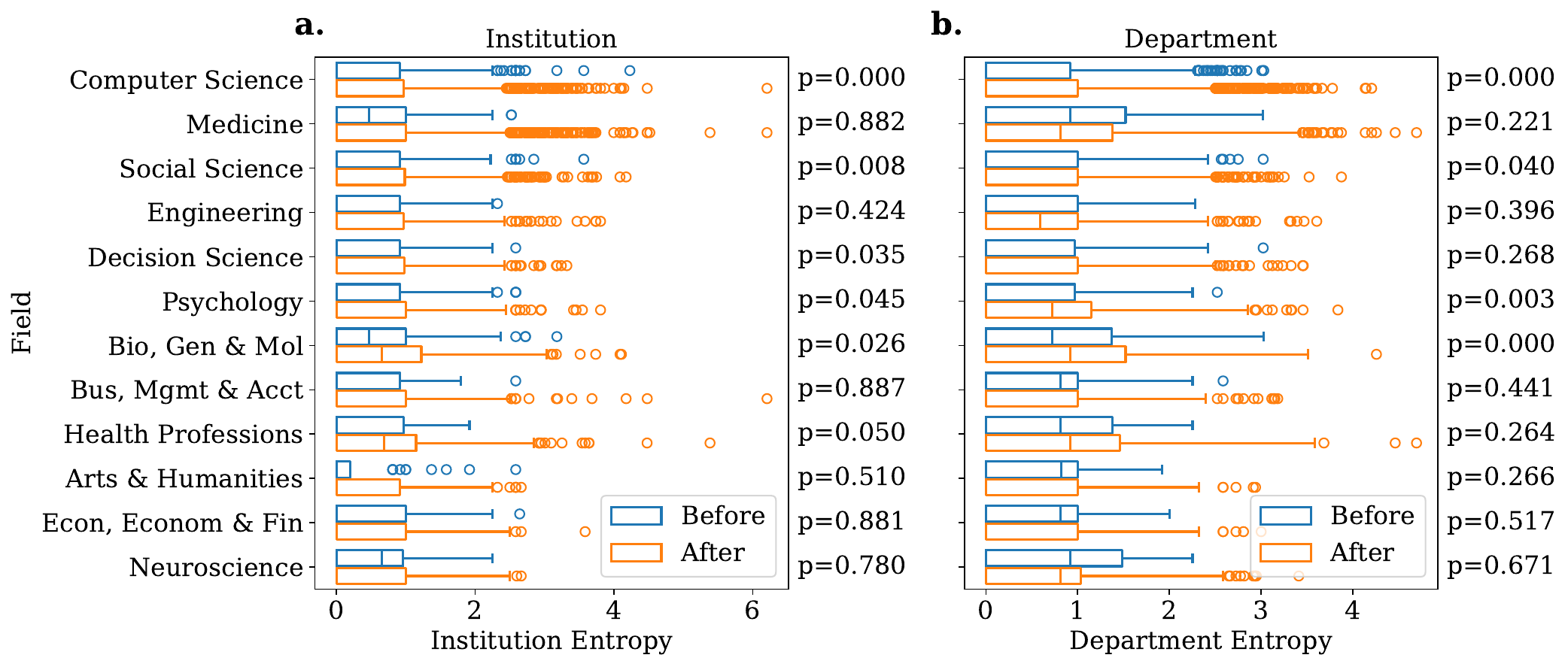}
  \caption{The entropy distribution in terms of each field before (Blue) and after (Red) the ChatGPT's release based on (a) authors' affiliated institutions, and (b) authors' affiliated departments. The x-axis shows the fields, and the y-axis denotes the Shannon Entropy. The Wilcoxon rank-sum test is conducted to compare the distribution of entropy before and after ChatGPT's release.}
  \label{fig:test_temporal}
\end{figure}

\begin{figure}[h!]
  \centering
  \hspace*{0.2cm}
  \includegraphics[width=0.98\textwidth]{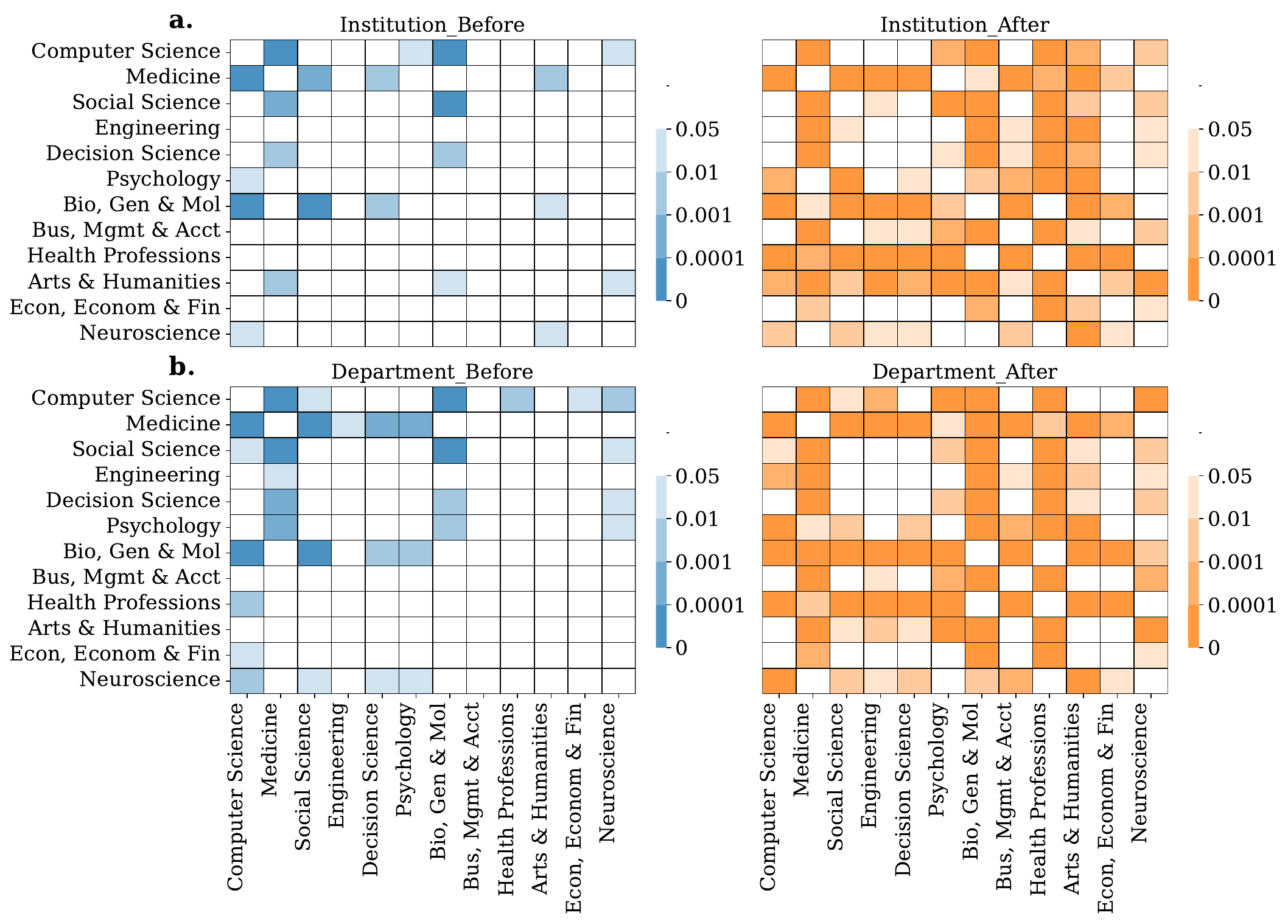}
  \caption{The Wilcoxon rank-sum test result across two fields before (left) and after (right) ChatGPT's release based on (a) authors' affiliated institutions, and (b) authors' affiliated departments. The lightest blue color implies that the two fields do not show significant differences in entropy, while the darker blue color represents the difference is more statistically significant.}
  \label{fig:test_fields}
\end{figure}

\autoref{fig:test_temporal} and \autoref{fig:test_fields} provide statistical support for the analysis presented in \autoref{collaboration}. Both figures employ the Wilcoxon rank-sum test, a non-parametric method ideal for comparing two groups of data that are either interval-scaled or not normally distributed, to quantify the statistical significance of observed differences in entropy. \autoref{fig:test_temporal} examines whether there is a significant difference in entropy before and after the release of ChatGPT, with subfigure (a) illustrating the entropy distribution and test results based on authors' institutions, and subfigure (b) presenting the same analysis for authors' departments. \autoref{fig:test_fields} investigates whether two fields display significant differences in entropy, with subfigure (a) showing the Wilcoxon rank-sum test results comparing entropy across two fields before and after ChatGPT's release based on authors' institutions, and subfigure (b) presenting the same comparison based on authors' departments.

\section{OpenAlex topic and field classification}
\label{openalex}

OpenAlex has identified 4,516 topics based on the publication-level classification system developed by \citet{waltman2012new}. The classification of topics for each paper is based on OpenAlex's proprietary model, which fine-tunes the multilingual BERT (mBERT) model for topic classification. The model's input comprises a paper's title, abstract, and citations, although only approximately half of the papers have usable abstracts. According to OpenAlex's performance report, when all information for a paper is available, the model achieves a (top K = 1) accuracy of 0.72. The (top K = 1) accuracy refers to the percentage of samples where the correct label appears as the top prediction. On average, their model achieves a (top K = 1) accuracy of 0.53 and a (top K = 3) accuracy of 0.73. OpenAlex provides up to three topics for each paper \cite{priem2022openalex}.

Subsequently, OpenAlex establishes a one-to-one relationship between these topics and higher-level fields. The fields are organized in a hierarchical structure, including subfields, fields, and domains, derived from Scopus's ASJC (All Science Journal Classification) structure. This matching process is conducted using an LLM and further verified by OpenAlex's annotators. As a result, each paper can be associated with up to three fields, corresponding to the three identified topics \cite{priem2022openalex}. 

While the classification performance for field classification is not explicitly reported, it is reasonable to assume that the accuracy for mapping the 4,516 topics to 26 fields could be significantly higher than the reported (top K = 1) accuracy for topic classification. This assumption is based on the fact that the field classification represents a narrowing-down process from a larger set of topics to a smaller set of fields \cite{priem2022openalex}.

\section{Top centrality entities}
\label{centrality}

\autoref{tab:top5_centrality_institutions} and \autoref{tab:top5_centrality_departments} present the top five entities based on degree centrality, betweenness centrality, closeness centrality, and eigenvector centrality, calculated from the authors' institutional and departmental affiliations, respectively. Each table lists the top five entities across three paper groups.

\begin{table}[htbp]
\centering
\caption{Top 5 entities based on authors' institution affiliations}
\begin{tabular}{p{2cm}p{3.8cm}p{3.8cm}p{4cm}}
\toprule
\textbf{Metric} & \textbf{LLM} & \textbf{ML} & \textbf{Non-LLM/ML} \\ 
\midrule
\textbf{Degree Centrality} & Stanford University (342) & Stanford University (667) & Centre National de la Recherche Scientifique (532) \\
& Tsinghua University (274) & University College London (612) & University College London (437) \\
& Peking University (263) & University of Oxford (561) & University of Toronto (393) \\
& University of Washington (261) & Harvard University (529) & University of California, Los Angeles (393) \\
& Harvard University (245) & University of Cambridge (523) & University of Oxford (376) \\
\hline
\textbf{Betweenness Centrality} & Instituto de Ciencias de La Construcción Eduardo Torroja (0.500) & Imperial College London (0.019) & King Saud University (0.017) \\
& Escuela Nacional de Sanidad (0.500) & University College London (0.017) & University of Toronto (0.016) \\
& Heilongjiang Academy of Sciences (0.500) & University of Oxford (0.016) & University of Oxford (0.014) \\
& China Academy of Space Technology (0.500) & Stanford University (0.015) & Chinese Academy of Sciences (0.013) \\
& Nahrain University (0.500) & New York University (0.014) & Centre National de la Recherche Scientifique (0.012) \\
\hline
\textbf{Closeness Centrality} & University College London (0.022) & Trilogi University (0.500) & African Studies Centre (0.500) \\
& University of Waterloo (0.021) & Universitas Bina Darma (0.500) & Institute for Development \& Economic Analysis (0.500) \\
& Yale University (0.017) & Nahrain University (0.500) & Semmes Murphey Foundation (0.500) \\
& Harvard University (0.017) & Heilongjiang Academy (0.500) & Kaleida Health (0.500) \\
& Peking University (0.016) & China Academy Space Tech. (0.500) & Buffalo General Medical Center (0.500) \\
\hline
\textbf{Eigenvector Centrality} & Shanghai Jiao Tong University (0.339) & Institute of High Energy Physics (0.503) & Institute of High Energy Physics (0.521) \\
& Shanghai Sixth People's Hospital (0.316) & Université Libre de Bruxelles (0.470) & Université Libre de Bruxelles (0.453) \\
& Stanford University (0.276) & UCLouvain (0.415) & Vrije Universiteit Brussel (0.402) \\
& Tsinghua University (0.254) & Vrije Universiteit Brussel (0.370) & UCLouvain (0.372) \\
& Chinese University of Hong Kong (0.252) & Ghent University Hospital (0.319) & University of Antwerp (0.329) \\
\bottomrule
\end{tabular}
\label{tab:top5_centrality_institutions}
\end{table}

\begin{table}[htbp]
\centering
\caption{Top 5 entities based on authors' department affiliations}
\begin{tabular}{p{2cm}p{3.8cm}p{3.8cm}p{4cm}}
\toprule
\textbf{Metric} & \textbf{LLM} & \textbf{ML} & \textbf{Non-LLM} \\ 
\midrule
\textbf{Degree Centrality} & Computer Science (1392) & Computer Science (3001) & Medicine (1948) \\
& Medicine (1019) & Medicine (2184) & Chemistry (1278) \\
& Computer Science and Engineering (611) & Engineering (1501) & Physics (1007) \\
& Artificial Intelligence (586) & Physics (1240) & Public Health (996) \\
& Engineering (385) & Computer Science and Engineering (1201) & Biology (879) \\
\hline
\textbf{Betweenness Centrality} & Dynamics, Logics and Inference for Biological Systems (0.500) & Computer Science (0.0761) & Medicine (0.041) \\
& Data and Knowledge Management (0.500) & Medicine (0.0372) & Chemistry (0.037) \\
& Computer Engineering and Interdisciplinary Major (0.500) & Engineering (0.0429) & Mechanical Engineering (0.030) \\
& Control \& Automation Engineering (0.500) & Mechanical Eng. (0.0298) & Engineering (0.029) \\
& Reasoning (0.500) & Physics (0.0273) & Physics (0.029) \\
\hline
\textbf{Closeness Centrality} & Computer Science (0.091) & Innovation \& Society (0.500) & Rail Systems Management (0.500) \\
& Medicine (0.043) & Responsible Innovation (0.500) & Aviation Management (0.500) \\
& Artificial Intelligence (0.036) & Data \& Knowledge Mgmt. (0.500) & Computing Science (0.500) \\
& Computer Science and Engineering (0.033) & Maritime Transportation Safety (0.500) & Control, Robotics, and Electrical Eng. (0.500) \\
& Education (0.031) & Political Science and International Relations (0.500) & Applied Analysis and Stochastics (0.500) \\
\hline
\textbf{Eigenvector Centrality} & Business (0.561) & Physics (0.7022) & Physics (0.698) \\
& Information Systems (0.311) & Astronomy (0.102) & Astronomy (0.102) \\
& Technology, Policy and Management (0.070) & Astrophysics (0.056) & Nuclear Physics (0.090) \\
& Management Studies (0.053) & Nuclear Physics (0.053) & Astrophysics (0.056) \\
& Marketing and Marketing Management (0.053) & Earth Sciences (0.022) & Astroparticle and Cosmology (0.050) \\
\bottomrule
\end{tabular}
\label{tab:top5_centrality_departments}
\end{table}

\end{appendices}


\bibliography{sn-bibliography}

\end{document}